\documentclass[aps,twocolumn,showpacs,groupedaddress,floatfix]{revtex4-1}%
\usepackage{graphicx}
\usepackage[dvips,usenames]{color}
\usepackage{txfonts}
\usepackage{units}
\usepackage{epsfig}
\newcommand{\uK}{\ensuremath{\mu{\mbox{K}}}}
\newcommand{\micron}{\ensuremath{\mu{\mbox{m}}}}

\newcommand{\half}{\ensuremath{\nicefrac{1}{2}}}

\begin{document}

\author{Y. A. Liao}
\author{M. Revelle}
\author{T. Paprotta}
\altaffiliation{Current address: Thorlabs Inc., Newton, NJ 07860}
\author{A. S. C. Rittner}
\altaffiliation{Current address: The Boston Consulting Group, 20095 Hamburg, Germany}
\author{Wenhui Li}
\altaffiliation{Permanent address: Centre for Quantum Technologies, National University of Singapore, 3 Science Drive 2, Singapore, 117543}
\author{G. B. Partridge}
\altaffiliation{Current address: Agilent Laboratories, Santa Clara, CA 95051}
\author{R. G. Hulet}
\affiliation{Department of Physics and Astronomy and Rice Quantum Institute,
Rice University, Houston, Texas 77005, USA}

\date{\today}

\title{Metastability in Spin-Polarized Fermi Gases}

\begin{abstract}
We study the role of particle transport and evaporation on the phase separation of an ultracold, spin-polarized atomic Fermi gas.  We show that the previously observed deformation of the superfluid paired core is a result of evaporative depolarization of the superfluid due to a combination of enhanced evaporation at the center of the trap and the inhibition of spin transport at the normal-superfluid phase boundary. These factors contribute to a nonequilibrium jump in the chemical potentials at the phase boundary. Once formed, the deformed state is highly metastable, persisting for times of up to 2 s.
\end{abstract}


\maketitle


The BCS theory of superconductivity is remarkably successful
in describing pairing of unpolarized spin-$\half$ particles. Pairing
in spin-polarized systems is much more complicated, however,
prompting speculation about exotic new pairing mechanisms
that began shortly after the development of
the BCS theory~\cite{FF, LO}, and continues until
today~\cite{Casalbuoni04, Sheehy07}.  Spin-polarization, or more generally,
imbalanced Fermi energies, arise in several physical situations including
certain superconductors that
support coexisting magnetic and superconducting order, color
superconductivity in quark matter, and in ultracold atomic gases
created with imbalanced spin populations. In 2006, a group at
Massachusetts Institute of Technology (MIT) ~\cite{Zwierlein06, Shin06} and our group at 
Rice University~\cite{Partridge06a,Partridge06b} discovered that
strongly-interacting spin-imbalanced trapped atomic gases undergo a
first-order phase separation between a fully-paired superfluid core
and lower density polarized regions.

There are significant qualitative and quantitative differences
between the MIT and Rice experiments.  The phase separation
in the case of the Rice experiment was characterized by strong
deformation of the paired core, in violation of the local-density
approximation (LDA).
In the LDA, the local chemical potentials
$\mu_\sigma(\textbf{r}) = \mu_\sigma - V(\textbf{r})$ depend
only on the trap potential $V(\textbf{r})$
and the spatially- uniform global chemical potentials $\mu_{\sigma}$.
Here, $\sigma = \uparrow, \downarrow$ designate the two states of a
pseudo-spin-$\half$ system.  The local densities are given by these
local chemical potentials and the equilibrium equation of state
of an infinite, spatially uniform system.
For a harmonically confined gas with an unpolarized central core,
the LDA implies a flattopped axial
spin-density (obtained by integrating the three-dimensional spin
density along the two radial coordinates)~\cite{DeSilva06, Haque06, Imambekov06}.
While this flattopped distribution was observed in the MIT experiment~\cite{Shin06},
in the Rice experiment the shape of the paired core was
significantly less elongated than $V(\textbf{r})$, resulting in a central
dip in the axial spin-density~\cite{Partridge06a,Partridge06b}.
Furthermore, pairing in the Rice experiment was much more robust
than in the MIT experiment, persisting to much larger population
imbalances.  This robust pairing is apparently in contradiction with
the Clogston-Chandrasekhar limit describing the break-down of pairing
when the difference between the chemical potentials of the two spin-states
exceeds the pairing gap~\cite{Chandrasekhar62, Clogston62, Lobo06, Shin08}.

\begin{figure}
\includegraphics[width=1.0\columnwidth]{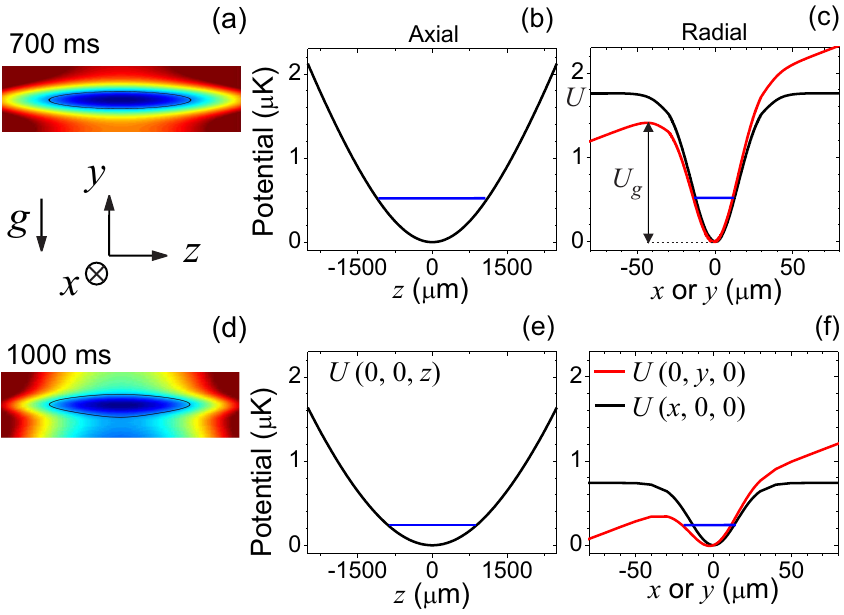}
\caption{(Color online) Plots of the hybrid magnetic-optical trap. Atoms
are trapped at the focus of a laser beam
propagating in the $z$ (axial) direction.  The laser
wavelength is 1.08 $\micron$ and the beam is focused with
a Gaussian beam waist ($1/e^2$ radius) of 30 $\micron$.
A residual
magnetic curvature contributes to the axial confinement with a
harmonic frequency of 3.8 Hz. (a)-(c) Optical trap depth $U = 1.68\,
\uK$, corresponding to $t = 700$ ms for the 1 s evaporation
trajectory. $U_g$ is the effective trap depth accounting for gravity.
The relative contribution of the magnetic curvature and
gravity are small. (d)-(f) Potential at the final optical trap depth
$U_f = 0.74\, \uK$ at $t = 1$ s, where the relatively strong
magnetic curvature has the effect of opening up a lip at $z=0$.
At this trap depth, the combined axial frequency, due to the optical
and magnetic forces is 4.7 Hz.
Values of k$_B T_F$ from Fig.~\ref{fig:evap_traj}(a) are indicated
by the horizontal blue lines.
\label{fig:trap} }
\end{figure}

Possible explanations for this discrepancy have focused on the
primary differences between the two experiments, which are trap
aspect ratio and particle number~\cite{Imambekov06, DeSilvaPRL06,
Haque07, Sensarma07, Tezuka08, Ku09, Baksmaty11}. In both cases, the
confining potential is approximately harmonic and elongated along
the cylindrical ($z$) axis, as shown in Fig.~\ref{fig:trap}(a).  For
the Rice experiment, however, the ratio of the radial to axial trap
frequencies was $\sim$30, while for MIT it was $\sim$5.  Also, the
total particle number in the Rice experiment was $\sim$$10^5$, while
for MIT it was $\sim$$5 \times 10^6$.  It was shown that while the
observed deformation was consistent with the effect of a strong
surface tension at the superfluid-normal
interface~\cite{Partridge06b, DeSilvaPRL06, Haque07}, the required
magnitude of the surface tension was inconsistent with detailed
calculations~\cite{Baur09, Diederix11}.  Furthermore, a recent
experiment is largely in agreement with the MIT results despite
having an aspect ratio and particle number that are similar to the
Rice experiment~\cite{Nascimbene09}.  A new mechanism has been
proposed ~\cite{Parish09} which has its origins in the inhibition of
thermal~\cite{Schaeybroeck07} and spin transport~\cite{Parish09}
across the phase boundaries coupled with an enhanced probability for
evaporation at the axial center of the trap.  These factors enable a
distribution that is out of chemical equilibrium, where the
difference in chemical potentials is depressed in the superfluid
phase at trap center relative to the polarized normal phases in the
wings. We have performed several experiments that strongly support
this conjecture. In addition, we find that, once produced, the
deformed state is remarkably metastable.

We produce imbalanced mixtures of $^6$Li atoms as before in the two
lowest energy hyperfine states ($F=\half, \,m_F=\half$) and
($F=\half, \,m_F=-\nicefrac{1}{2}$), designated as states
$|$$\uparrow\rangle$ and $|$$\downarrow\rangle$,
respectively~\cite{Partridge06a,Partridge06b}. A bias magnetic field
is tuned to 834 G, corresponding to the unitary limit of the $^6$Li
Feshbach resonance.
The atoms are confined in a hybrid
optical-magnetic trap formed by a single focused laser beam
propagating along the direction of the bias field (axial direction).
Radial confinement is produced by the Gaussian intensity profile
of the laser beam, while axial confinement arises from the combination
of the Lorentzian axial profile of the laser beam and the residual magnetic
curvature (confining) from the slightly non-Helmholtz configuration of the
magnetic bias coils.
This combination, depicted in
Fig.~\ref{fig:trap}(a), results in an aspect ratio of 90$\sim$100 with
an approximately isotropic trap depth when the optical potential
dominates. Evaporative cooling is effected by lowering the laser
power, such that the trap depth, as well as the trap aspect ratio,
is gradually reduced.  At sufficiently low optical power the
magnetic curvature dominates the axial confinement.
When this happens, the trap depth becomes anisotropic,
with the depth being largest along the axial (magnetic curvature) direction.
Thus, a ``lip'' of minimum trap depth is formed in the radial direction
at the axial trap center ($z=0$),
as shown in Fig.~\ref{fig:trap}(d).
Furthermore, gravity reduces the trap depth in the direction
pointing downward in the lab (along $-y$).

\begin{figure}[b]
\includegraphics[width=1.0\columnwidth]{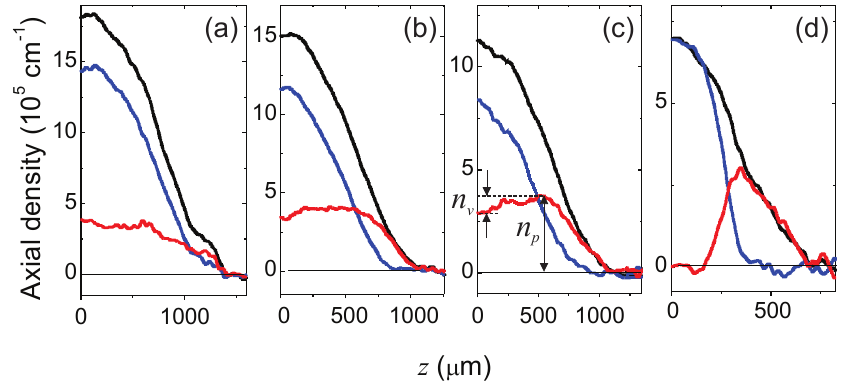}
\caption{(Color online) Axial densities taken at various times $t$ along the
evaporation trajectory.  The global polarizations $P$ are:
(a) 600 ms, $P=0.14$,
(b) 700 ms, $P=0.21$,
(c) 740 ms, $P=0.28$,
(d) 1 s, $P=0.18$.  The variation in $P$ is a result of
shot-to-shot variations, as each image requires the trap to be reloaded
and evaporated to the specified $t$.
The upper (black) curves correspond to
the majority state ($|$$\uparrow\rangle$), the middle (blue) curves to the minority state
($|$$\downarrow\rangle$), and the axial spin densities 
($|$$\uparrow\rangle - |$$\downarrow\rangle$) are given by the lower (red) curves.
\label{fig:axial_densities}}
\end{figure}

Sequential absorption images~\cite{Partridge06b} record the column
density distributions of the trapped atoms for each state.  Figure
\ref{fig:axial_densities} shows representative axial density
profiles, obtained by integrating the column density images along
the remaining radial coordinate, for images recorded at various
times along the evaporation trajectory used in our previous studies.
For this trajectory, the trap depth was reduced exponentially as
$e^{-t/\tau}$ from its initial value $U_i = 160\, \uK$ to a final
value $U_f = 0.74\, \uK$ in a total time $t_{tot} = 1$ s, and with
an exponential time constant $\tau = 200$ ms.  Deformation is evidenced
by a dip in the axial spin density, which begins to develop at approximately
$t = $ 700 ms [Fig.~\ref{fig:axial_densities}(b)],
corresponding to a temperature $T \simeq 0.2 \,T_F$, where $T_F$ is the
Fermi temperature of a noninteracting trapped gas of $|$$\uparrow\rangle$ atoms.
The deformation, characterized by the parameter
$\alpha = n_v / n_p$ [see Fig.~\ref{fig:axial_densities}(c)], increases as the evaporation progresses and
is maximum at $t_{tot}$, where $T \simeq 0.06 \,T_F$ is at its minimum.

\begin{figure}
\includegraphics[width=1.0\columnwidth]{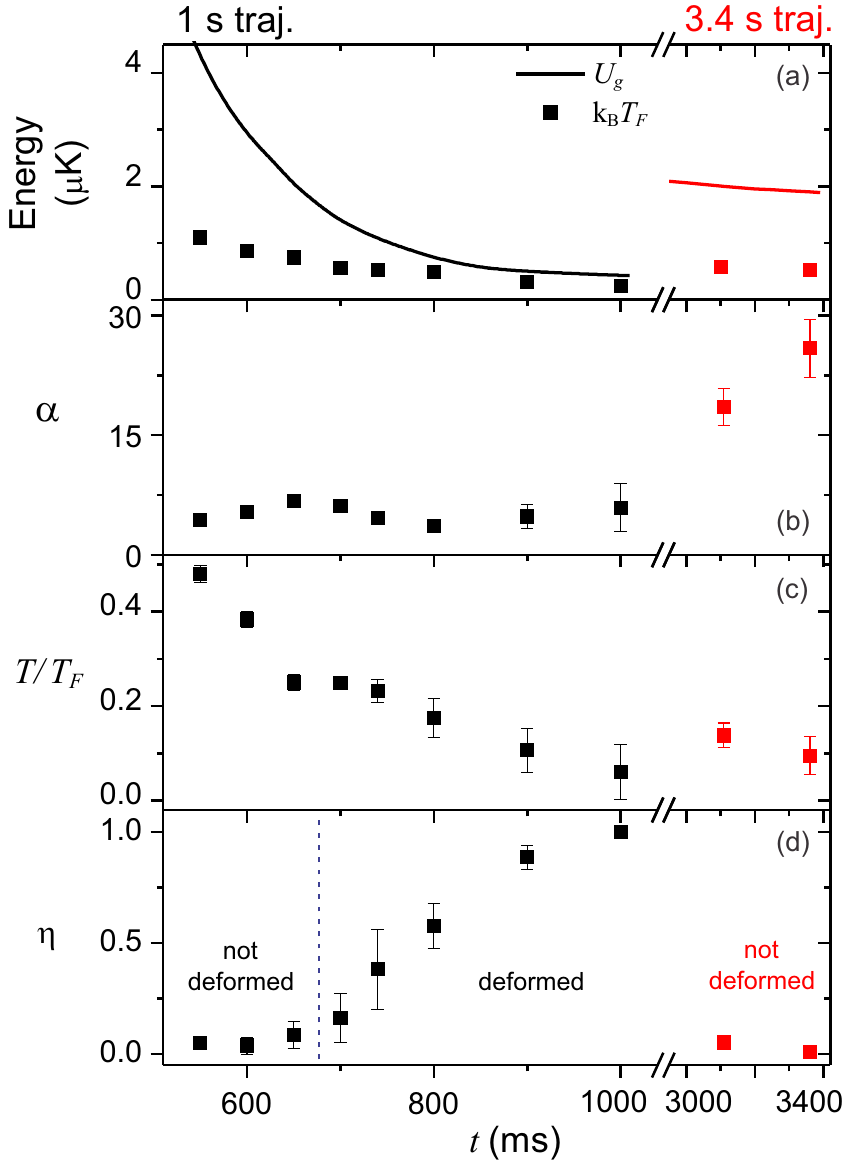}
\caption{(Color online) Parameters extracted from the axial densities for
both the (aggressive) 1 s and (gentle) 3.4 s trajectories. $T$ is determined
from the mean energy $E$ of unpolarized ($P=0$) distributions which are separately evaporated using the same trajectory. $E$ for each distribution is obtained from its mean-squared radius via the virial theorem~\cite{Luo09}. The $E$ vs~$T$ calibration is given in Ref.~\cite{Hu10} and is based on the experimental data of Ref.~\cite{Nascimbene10}.
$\eta = (U_g - \epsilon_p)/\textrm{k}_BT$, where
$\epsilon_p = \half\,m\omega_z^2R^2$, $\omega_z$ is the axial trap
frequency, and $R$ is the axial radius where the density of the majority
state ($\uparrow$) goes to zero.  The deformation parameter is defined as
$\alpha = n_v / n_p$, as depicted in Fig.~\ref{fig:axial_densities}(c).
The error bars are mainly statistical uncertainty from the average of $\sim$6
shots of various values of $P$ at each value of $t$.  The vertical dashed
line in (d) indicates the onset of deformation.
\label{fig:evap_traj} }
\end{figure}

\begin{figure}
\includegraphics[width=1.0\columnwidth]{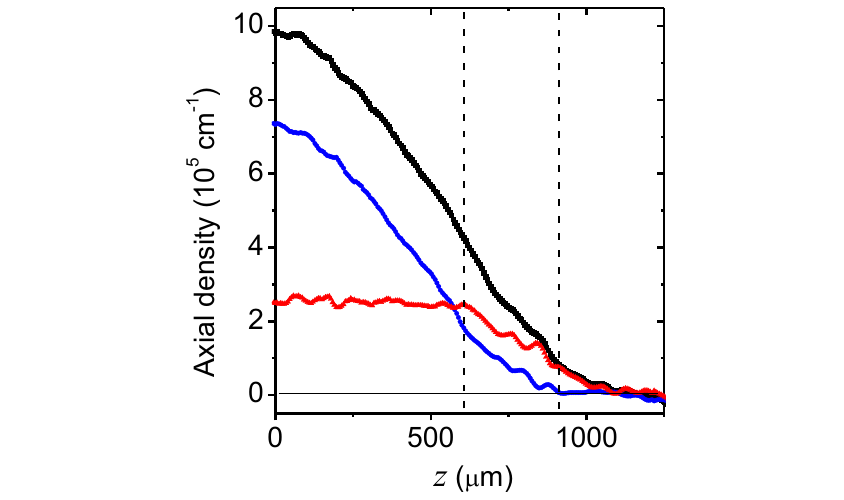}
\caption{(Color online) Axial densities for the 3.4 s trajectory at
$t = 3.4$ s, with $P = 0.24$.
Curve designations are the same as in Fig.~\ref{fig:axial_densities}.
The dashed vertical lines indicate the location of phase boundaries.
The flattopped axial spin density is consistent with the LDA, even
though the aspect ratio of the trap potential is $\sim$96.
\label{fig:LDAden}}
\end{figure}

Figure \ref{fig:evap_traj} shows the progression of several relevant parameters
during evaporation.  $U_g$, shown in Fig.~\ref{fig:evap_traj}(a), is the trap
depth including gravity, as defined in Fig.~\ref{fig:trap}(c).  Also shown in Fig.~\ref{fig:evap_traj} are parameters extracted from
a less aggressive evaporation trajectory for which $U_i = 160\, \uK$
as before, but now with $U_f = 2.2\, \uK$, $t_{tot} = 3.4$ s, and $\tau =
500$ ms.  This trajectory is designed to be similar to the final part
of the trajectory used in
Ref.~\cite{Nascimbene09}, where no deformation was observed. Figure
~\ref{fig:evap_traj}(b) shows the value of $\eta = (U_g -
\epsilon_p)/\textrm{k}_BT$, where $\epsilon_p = \half\,m\omega_z^2R^2$, $\omega_z$
is the axial trap frequency, and $R$ is the axial radius where the density
of the majority state ($\uparrow$) goes to zero.  The value of $\eta$ is an
approximate measure of
the closeness of the chemical potential to the trap lip, and hence
is related to the rate of evaporation.  This quantity is
significantly larger for the ``gentle" 3.4 s trajectory as compared to the
``aggressive" 1 s trajectory, indicating a much slower rate of evaporation.
Nonetheless, even though the 3.4 s trajectory is not
as deep or as aggressive, the final temperature of $\sim$0.09 $T_F$ is
similar to that achieved with the 1 s trajectory.  Furthermore, as
shown in Fig.~\ref{fig:LDAden}, the axial spin density at the end of the
trajectory is flattopped, indicating that there is no deformation even
though the trap aspect ratio at the end of evaporation is highly
elongated (aspect ratio of $\sim$96).  Deformation is prevented in
the 3.4 s trajectory by its higher final trap depth, which both
reduces the rate of evaporation and minimizes the lip at $z=0$.

\begin{figure}
\includegraphics[width=1.0\columnwidth]{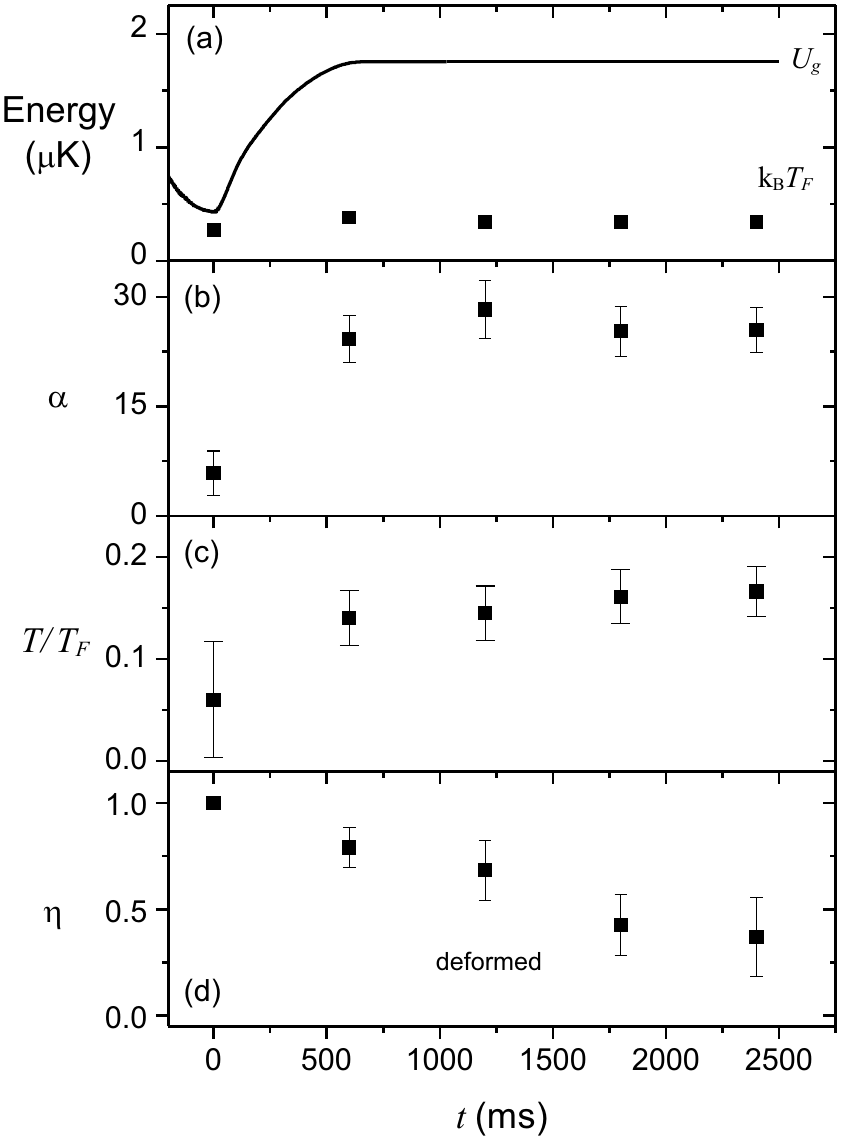}
\caption{Recompression of the trap following the 1 s evaporation.
Here, the 1 s evaporation finishes at $t = 0$ and is followed by a
slow recompression of the trap over the next 0.6 s. Even though
the final value of $U_g$ is similar to that shown for the 3.4 s
trajectory in Fig.~\ref{fig:evap_traj} the recompressed values
of $T_F$ are lower due to smaller overall numbers
($\sim$$3 \times 10^4$ vs~$\sim$$10^5$).
\label{fig:recompress} }
\end{figure}

To determine whether the deformed state is only dynamically stable,
existing only during rapid anisotropic evaporation, or rather is a
metastable state, we ramped the trap depth up over a time period of 600 ms
following evaporation,
as shown in Fig.~\ref{fig:recompress}(a). This serves to significantly
suppress the rate of evaporation, as can be seen from the nearly constant
value of $T_F$ in
Fig.~\ref{fig:recompress}(a) and the large values of $\eta$ in
Fig.~\ref{fig:recompress}(b).
Nonetheless, Fig.~\ref{fig:recompress}(d) shows that the deformation
$\alpha$ remains for more than 2 s following trap recompression.  The degree
of deformation is seen to decrease following recompression, roughly on the
same timescale of an observed rise in the temperature.

\begin{figure}
\includegraphics[width=1.0\columnwidth]{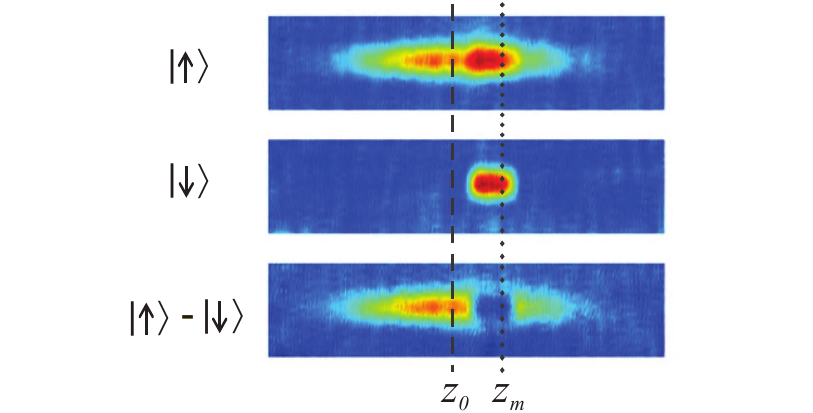}
\caption{(Color online) Column density images for an axial displacement of the
center of magnetic curvature located at $z_m = 0$ and indicated by the
dotted vertical line, from the focus of the optical trap beam
located at $z = -1$ mm. At this trap depth ($U = 0.65\, \uK$), the center of the
combined potential, indicated by the dashed vertical line, is located at
$z_0 = -210\, \micron$.  The trap beam waist is $26\, \micron$ and
the residual magnetic curvature is 6.4 Hz for these data.
The image size is $1653\, \micron \times 100\, \micron$ and $P = 0.63$. The uncertainty in $z_0$ and $z_m$ is 20 $\mu$m due to uncertainties in the optical trap parameters.
\label{fig:offcenter} }
\end{figure}

Another measure of the effect of the lip in the potential may be
obtained by axially displacing the center of the magnetic curvature
with respect to the focus of the optical trap laser beam, as
depicted in Fig.~\ref{fig:offcenter}. Since the lip is located at
the minimum of the magnetic curvature ($z = z_m$), its position no
longer coincides with the overall minimum of the combined
magnetic-optical potential, indicated by $z_0$ in
Fig.~\ref{fig:offcenter}.  Figure \ref{fig:offcenter} shows that
while the unpaired atoms, given by the distribution of
$|$$\uparrow\rangle-|$$\downarrow\rangle$, reside near the center of
the combined potential ($z_0$), the paired core, given by the
$|$$\downarrow\rangle$ distribution, displaces towards the lip at $z
= z_m$ where evaporative cooling has maximum effect. This
observation provides a graphic illustration of the lack of
equilibration between the superfluid core and the normal phases.

We have presented a series of measurements that are consistent with a model of evaporative depolarization \cite{Parish09}. In this model, the chemical potential of the majority species is depleted by preferential evaporation in the vicinity of a lip in the elongated trap potential. Because of the inhibition of spin transport, for sufficiently fast evaporation the chemical potential difference can be smaller in the paired core than in the polarized wings. This helps to stabilize the superfluid phase beyond the Clogston-Chandrasekhar limit, as was observed experimentally \cite{Partridge06a,Partridge06b}. The absence of a spatial variation in the spin-density of the superfluid phase \cite{Partridge06b} shows that the gradient in the chemical potential difference is not smoothly varying, but rather that it undergoes a sudden jump at the phase boundary. This indicates that the interface, not the bulk superfluid, is the dominant obstacle to spin transport. Slow relaxation of a nonequilibrium spin distribution in a spin-imbalanced phase separated gas was also reported in Ref.~ \cite{Sommer11b}. The observed relaxation time scales in that experiment are consistent with ours but the relative importance of the interface compared with the bulk superfluid could not be determined. (A previous experiment also reported slow diffusion but it was conducted with a balanced spin mixture, and hence without phase boundaries \cite{Sommer11a}). In our experiment, the jump in the local chemical potential difference at the phase boundary strongly deforms the shape of the superfluid-normal interface, making it much less curved than it would be at equilibrium. This nonequilibrium distribution is remarkably metastable, consistent with the calculations presented in Ref.~\cite{Baksmaty11}, in which they find both LDA-like solutions as well as nearly degenerate LDA-violating ones whose density distributions closely resemble those that we observe.


We thank David Huse and Henk Stoof for valuable discussions. This
work was supported under ARO Grant No. W911NF-07-1-0464 with funds from
the DARPA OLE program, and by the NSF, the ONR, and the Welch
Foundation (Grant No. C-1133).

\bibliographystyle{apsrev4-1}

\begin{thebibliography}{31}%
\makeatletter
\providecommand \@ifxundefined [1]{%
 \@ifx{#1\undefined}
}%
\providecommand \@ifnum [1]{%
 \ifnum #1\expandafter \@firstoftwo
 \else \expandafter \@secondoftwo
 \fi
}%
\providecommand \@ifx [1]{%
 \ifx #1\expandafter \@firstoftwo
 \else \expandafter \@secondoftwo
 \fi
}%
\providecommand \natexlab [1]{#1}%
\providecommand \enquote  [1]{``#1''}%
\providecommand \bibnamefont  [1]{#1}%
\providecommand \bibfnamefont [1]{#1}%
\providecommand \citenamefont [1]{#1}%
\providecommand \href@noop [0]{\@secondoftwo}%
\providecommand \href [0]{\begingroup \@sanitize@url \@href}%
\providecommand \@href[1]{\@@startlink{#1}\@@href}%
\providecommand \@@href[1]{\endgroup#1\@@endlink}%
\providecommand \@sanitize@url [0]{\catcode `\\12\catcode `\$12\catcode
  `\&12\catcode `\#12\catcode `\^12\catcode `\_12\catcode `\%12\relax}%
\providecommand \@@startlink[1]{}%
\providecommand \@@endlink[0]{}%
\providecommand \url  [0]{\begingroup\@sanitize@url \@url }%
\providecommand \@url [1]{\endgroup\@href {#1}{\urlprefix }}%
\providecommand \urlprefix  [0]{URL }%
\providecommand \Eprint [0]{\href }%
\providecommand \doibase [0]{http://dx.doi.org/}%
\providecommand \selectlanguage [0]{\@gobble}%
\providecommand \bibinfo  [0]{\@secondoftwo}%
\providecommand \bibfield  [0]{\@secondoftwo}%
\providecommand \translation [1]{[#1]}%
\providecommand \BibitemOpen [0]{}%
\providecommand \bibitemStop [0]{}%
\providecommand \bibitemNoStop [0]{.\EOS\space}%
\providecommand \EOS [0]{\spacefactor3000\relax}%
\providecommand \BibitemShut  [1]{\csname bibitem#1\endcsname}%
\let\auto@bib@innerbib\@empty
\bibitem [{\citenamefont {Fulde}\ and\ \citenamefont {Ferrell}(1964)}]{FF}%
  \BibitemOpen
  \bibfield  {author} {\bibinfo {author} {\bibfnamefont {P.}~\bibnamefont
  {Fulde}}\ and\ \bibinfo {author} {\bibfnamefont {R.~A.}\ \bibnamefont
  {Ferrell}},\ }\href {\doibase 10.1103/PhysRev.135.A550} {\bibfield  {journal}
  {\bibinfo  {journal} {Phys. Rev.}\ }\textbf {\bibinfo {volume} {135}},\
  \bibinfo {pages} {A550} (\bibinfo {year} {1964})}\BibitemShut {NoStop}%
\bibitem [{\citenamefont {Larkin}\ and\ \citenamefont
  {Ovchinnikov}(1964)}]{LO}%
  \BibitemOpen
  \bibfield  {author} {\bibinfo {author} {\bibfnamefont {A.~I.}\ \bibnamefont
  {Larkin}}\ and\ \bibinfo {author} {\bibfnamefont {Y.~N.}\ \bibnamefont
  {Ovchinnikov}},\ }\href@noop {} {\bibfield  {journal} {\bibinfo  {journal}
  {Zh. Eksp. Teor. Fiz.}\ }\textbf {\bibinfo {volume} {47}},\ \bibinfo {pages}
  {1136} (\bibinfo {year} {1964})},\ \bibinfo {note} {[Sov. Phys. JETP \textbf{20}, 762
  (1965)}\BibitemShut {NoStop}%
\bibitem [{\citenamefont {Casalbuoni}\ and\ \citenamefont
  {Nardulli}(2004)}]{Casalbuoni04}%
  \BibitemOpen
  \bibfield  {author} {\bibinfo {author} {\bibfnamefont {R.}~\bibnamefont
  {Casalbuoni}}\ and\ \bibinfo {author} {\bibfnamefont {G.}~\bibnamefont
  {Nardulli}},\ }\href {\doibase 10.1103/RevModPhys.76.263} {\bibfield
  {journal} {\bibinfo  {journal} {Rev. Mod. Phys.}\ }\textbf {\bibinfo {volume}
  {76}},\ \bibinfo {pages} {263} (\bibinfo {year} {2004})}\BibitemShut
  {NoStop}%
\bibitem [{\citenamefont {Sheehy}\ and\ \citenamefont
  {Radzihovsky}(2007)}]{Sheehy07}%
  \BibitemOpen
  \bibfield  {author} {\bibinfo {author} {\bibfnamefont {D.~E.}\ \bibnamefont
  {Sheehy}}\ and\ \bibinfo {author} {\bibfnamefont {L.}~\bibnamefont
  {Radzihovsky}},\ }\href {\doibase DOI: 10.1016/j.aop.2006.09.009} {\bibfield
  {journal} {\bibinfo  {journal} {Annals of Physics}\ }\textbf {\bibinfo
  {volume} {322}},\ \bibinfo {pages} {1790} (\bibinfo {year}
  {2007})}\BibitemShut {NoStop}%
\bibitem [{\citenamefont {Zwierlein}\ \emph {et~al.}(2006)\citenamefont
  {Zwierlein}, \citenamefont {Schirotzek}, \citenamefont {Schunck},\ and\ \citenamefont {Ketterle}}]{Zwierlein06}%
  \BibitemOpen
  \bibfield  {author} {\bibinfo {author} {\bibfnamefont {M.~W.}\ \bibnamefont
  {Zwierlein}}, \bibinfo {author} {\bibfnamefont {A.}~\bibnamefont
  {Schirotzek}}, \bibinfo {author} {\bibfnamefont {C.~H.}\ \bibnamefont
  {Schunck}}, and\ \bibinfo {author} {\bibfnamefont {W.}~\bibnamefont
  {Ketterle}},\ }\href {\doibase 10.1126/science.1122318} {\bibfield  {journal}
  {\bibinfo  {journal} {Science}\ }\textbf {\bibinfo {volume} {311}},\ \bibinfo
  {pages} {492} (\bibinfo {year} {2006})}\BibitemShut {NoStop}%
\bibitem [{\citenamefont {Shin}\ \emph {et~al.}(2006)\citenamefont {Shin},
  \citenamefont {Zwierlein}, \citenamefont {Schunck}, \citenamefont
  {Schirotzek},\ and\ \citenamefont {Ketterle}}]{Shin06}%
  \BibitemOpen
  \bibfield  {author} {\bibinfo {author} {\bibfnamefont {Y.}~\bibnamefont
  {Shin}}, \bibinfo {author} {\bibfnamefont {M.~W.}\ \bibnamefont {Zwierlein}},
  \bibinfo {author} {\bibfnamefont {C.~H.}\ \bibnamefont {Schunck}}, \bibinfo
  {author} {\bibfnamefont {A.}~\bibnamefont {Schirotzek}}, and\ \bibinfo
  {author} {\bibfnamefont {W.}~\bibnamefont {Ketterle}},\ }\href {\doibase
  10.1103/PhysRevLett.97.030401} {\bibfield  {journal} {\bibinfo  {journal}
  {Phys. Rev. Lett.}\ }\textbf {\bibinfo {volume} {97}},\ \bibinfo {pages}
  {030401} (\bibinfo {year} {2006})}\BibitemShut {NoStop}%
\bibitem [{\citenamefont {Partridge}\ \emph
  {et~al.}(2006{\natexlab{a}})\citenamefont {Partridge}, \citenamefont {Li},
  \citenamefont {Kamar}, \citenamefont {Liao},\ and\ \citenamefont
  {Hulet}}]{Partridge06a}%
  \BibitemOpen
  \bibfield  {author} {\bibinfo {author} {\bibfnamefont {G.~B.}\ \bibnamefont
  {Partridge}}, \bibinfo {author} {\bibfnamefont {W.}~\bibnamefont {Li}},
  \bibinfo {author} {\bibfnamefont {R.~I.}\ \bibnamefont {Kamar}}, \bibinfo
  {author} {\bibfnamefont {Y.~A.}\ \bibnamefont {Liao}}, and\ \bibinfo
  {author} {\bibfnamefont {R.~G.}\ \bibnamefont {Hulet}},\ }\href {\doibase
  10.1126/science.1122876} {\bibfield  {journal} {\bibinfo  {journal}
  {Science}\ }\textbf {\bibinfo {volume} {311}},\ \bibinfo {pages} {503}
  (\bibinfo {year} {2006}{\natexlab{a}})}\BibitemShut {NoStop}%
\bibitem [{\citenamefont {Partridge}\ \emph
  {et~al.}(2006{\natexlab{b}})\citenamefont {Partridge}, \citenamefont {Li},
  \citenamefont {Liao}, \citenamefont {Hulet}, \citenamefont {Haque},\ and\
  \citenamefont {Stoof}}]{Partridge06b}%
  \BibitemOpen
  \bibfield  {author} {\bibinfo {author} {\bibfnamefont {G.~B.}\ \bibnamefont
  {Partridge}}, \bibinfo {author} {\bibfnamefont {W.}~\bibnamefont {Li}},
  \bibinfo {author} {\bibfnamefont {Y.~A.}\ \bibnamefont {Liao}}, \bibinfo
  {author} {\bibfnamefont {R.~G.}\ \bibnamefont {Hulet}}, \bibinfo {author}
  {\bibfnamefont {M.}~\bibnamefont {Haque}}, and\ \bibinfo {author}
  {\bibfnamefont {H.~T.~C.}\ \bibnamefont {Stoof}},\ }\href {\doibase
  10.1103/PhysRevLett.97.190407} {\bibfield  {journal} {\bibinfo  {journal}
  {Phys. Rev. Lett.}\ }\textbf {\bibinfo {volume} {97}},\ \bibinfo {pages}
  {190407} (\bibinfo {year} {2006}{\natexlab{b}})}\BibitemShut {NoStop}%
\bibitem [{\citenamefont {De~Silva}\ and\ \citenamefont
  {Mueller}(2006{\natexlab{a}})}]{DeSilva06}%
  \BibitemOpen
  \bibfield  {author} {\bibinfo {author} {\bibfnamefont {T.~N.}\ \bibnamefont
  {De~Silva}}\ and\ \bibinfo {author} {\bibfnamefont {E.~J.}\ \bibnamefont
  {Mueller}},\ }\href {\doibase 10.1103/PhysRevA.73.051602} {\bibfield
  {journal} {\bibinfo  {journal} {Phys. Rev. A}\ }\textbf {\bibinfo {volume}
  {73}},\ \bibinfo {pages} {051602} (\bibinfo {year}
  {2006}{\natexlab{a}})}\BibitemShut {NoStop}%
\bibitem [{\citenamefont {Haque}\ and\ \citenamefont {Stoof}(2006)}]{Haque06}%
  \BibitemOpen
  \bibfield  {author} {\bibinfo {author} {\bibfnamefont {M.}~\bibnamefont
  {Haque}}\ and\ \bibinfo {author} {\bibfnamefont {H.~T.~C.}\ \bibnamefont
  {Stoof}},\ }\href {\doibase 10.1103/PhysRevA.74.011602} {\bibfield  {journal}
  {\bibinfo  {journal} {Phys. Rev. A}\ }\textbf {\bibinfo {volume} {74}},\
  \bibinfo {pages} {011602} (\bibinfo {year} {2006})}\BibitemShut {NoStop}%
\bibitem [{\citenamefont {Imambekov}\ \emph {et~al.}(2006)\citenamefont
  {Imambekov}, \citenamefont {Bolech}, \citenamefont {Lukin},\ and\
  \citenamefont {Demler}}]{Imambekov06}%
  \BibitemOpen
  \bibfield  {author} {\bibinfo {author} {\bibfnamefont {A.}~\bibnamefont
  {Imambekov}}, \bibinfo {author} {\bibfnamefont {C.~J.}\ \bibnamefont
  {Bolech}}, \bibinfo {author} {\bibfnamefont {M.}~\bibnamefont {Lukin}}, and\ \bibinfo {author} {\bibfnamefont {E.}~\bibnamefont {Demler}},\ }\href
  {\doibase 10.1103/PhysRevA.74.053626} {\bibfield  {journal} {\bibinfo
  {journal} {Phys. Rev. A}\ }\textbf {\bibinfo {volume} {74}},\ \bibinfo
  {pages} {053626} (\bibinfo {year} {2006})}\BibitemShut {NoStop}%
\bibitem [{\citenamefont {Chandrasekhar}(1962)}]{Chandrasekhar62}%
  \BibitemOpen
  \bibfield  {author} {\bibinfo {author} {\bibfnamefont {B.~S.}\ \bibnamefont
  {Chandrasekhar}},\ }\href@noop {} {\bibfield  {journal} {\bibinfo  {journal}
  {Appl. Phys. Lett.}\ }\textbf {\bibinfo {volume} {1}},\ \bibinfo {pages} {7}
  (\bibinfo {year} {1962})}\BibitemShut {NoStop}%
\bibitem [{\citenamefont {Clogston}(1962)}]{Clogston62}%
  \BibitemOpen
  \bibfield  {author} {\bibinfo {author} {\bibfnamefont {A.~M.}\ \bibnamefont
  {Clogston}},\ }\href {\doibase 10.1103/PhysRevLett.9.266} {\bibfield
  {journal} {\bibinfo  {journal} {Phys. Rev. Lett.}\ }\textbf {\bibinfo
  {volume} {9}},\ \bibinfo {pages} {266} (\bibinfo {year} {1962})}\BibitemShut
  {NoStop}%
\bibitem [{\citenamefont {Lobo}\ \emph {et~al.}(2006)\citenamefont {Lobo},
  \citenamefont {Recati}, \citenamefont {Giorgini},\ and\ \citenamefont
  {Stringari}}]{Lobo06}%
  \BibitemOpen
  \bibfield  {author} {\bibinfo {author} {\bibfnamefont {C.}~\bibnamefont
  {Lobo}}, \bibinfo {author} {\bibfnamefont {A.}~\bibnamefont {Recati}},
  \bibinfo {author} {\bibfnamefont {S.}~\bibnamefont {Giorgini}}, and\
  \bibinfo {author} {\bibfnamefont {S.}~\bibnamefont {Stringari}},\ }\href
  {\doibase 10.1103/PhysRevLett.97.200403} {\bibfield  {journal} {\bibinfo
  {journal} {Phys. Rev. Lett.}\ }\textbf {\bibinfo {volume} {97}},\ \bibinfo
  {pages} {200403} (\bibinfo {year} {2006})}\BibitemShut {NoStop}%
\bibitem [{\citenamefont {Shin}\ \emph {et~al.}(2008)\citenamefont {Shin},
  \citenamefont {Schunck}, \citenamefont {Schirotzek},\ and\ \citenamefont
  {Ketterle}}]{Shin08}%
  \BibitemOpen
  \bibfield  {author} {\bibinfo {author} {\bibfnamefont {Y.}~\bibnamefont
  {Shin}}, \bibinfo {author} {\bibfnamefont {C.~H.}\ \bibnamefont {Schunck}},
  \bibinfo {author} {\bibfnamefont {A.}~\bibnamefont {Schirotzek}}, and\
  \bibinfo {author} {\bibfnamefont {W.}~\bibnamefont {Ketterle}},\ }\href
  {\doibase {10.1038/nature06473}} {\bibfield  {journal} {\bibinfo  {journal}
  {Nature}\ }\textbf {\bibinfo {volume} {451}},\ \bibinfo {pages} {689}
  (\bibinfo {year} {2008})}\BibitemShut {NoStop}
\bibitem [{\citenamefont {De~Silva}\ and\ \citenamefont
  {Mueller}(2006{\natexlab{b}})}]{DeSilvaPRL06}%
  \BibitemOpen
  \bibfield  {author} {\bibinfo {author} {\bibfnamefont {T.~N.}\ \bibnamefont
  {De~Silva}}\ and\ \bibinfo {author} {\bibfnamefont {E.~J.}\ \bibnamefont
  {Mueller}},\ }\href {\doibase 10.1103/PhysRevLett.97.070402} {\bibfield
  {journal} {\bibinfo  {journal} {Phys. Rev. Lett.}\ }\textbf {\bibinfo
  {volume} {97}},\ \bibinfo {pages} {070402} (\bibinfo {year}
  {2006}{\natexlab{b}})}\BibitemShut {NoStop}%
\bibitem [{\citenamefont {Haque}\ and\ \citenamefont {Stoof}(2007)}]{Haque07}%
  \BibitemOpen
  \bibfield  {author} {\bibinfo {author} {\bibfnamefont {M.}~\bibnamefont
  {Haque}}\ and\ \bibinfo {author} {\bibfnamefont {H.~T.~C.}\ \bibnamefont
  {Stoof}},\ }\href {\doibase 10.1103/PhysRevLett.98.260406} {\bibfield
  {journal} {\bibinfo  {journal} {Phys. Rev. Lett.}\ }\textbf {\bibinfo
  {volume} {98}},\ \bibinfo {pages} {260406} (\bibinfo {year}
  {2007})}\BibitemShut {NoStop}%
\bibitem [{\citenamefont {Sensarma}\ \emph {et~al.}(2007)\citenamefont
  {Sensarma}, \citenamefont {Schneider}, \citenamefont {Diener},\ and\
  \citenamefont {Randeria}}]{Sensarma07}%
  \BibitemOpen
  \bibfield  {author} {\bibinfo {author} {\bibfnamefont {R.}~\bibnamefont
  {Sensarma}}, \bibinfo {author} {\bibfnamefont {W.}~\bibnamefont {Schneider}},
  \bibinfo {author} {\bibfnamefont {R.~B.}\ \bibnamefont {Diener}}, and\
  \bibinfo {author} {\bibfnamefont {M.}~\bibnamefont {Randeria}},\ }\href@noop
  {} {\bibfield  {journal} {\bibinfo  {journal} {arXiv:0706.1741}\ } (\bibinfo
  {year} {2007})}\BibitemShut {NoStop}%
\bibitem [{\citenamefont {Tezuka}\ \emph {et~al.}(2010)\citenamefont {Tezuka},
  \citenamefont {Yanase},\ and\ \citenamefont {Ueda}}]{Tezuka08}%
  \BibitemOpen
  \bibfield  {author} {\bibinfo {author} {\bibfnamefont {M.}~\bibnamefont
  {Tezuka}}, \bibinfo {author} {\bibfnamefont {Y.}~\bibnamefont {Yanase}},
  and\ \bibinfo {author} {\bibfnamefont {M.}~\bibnamefont {Ueda}},\ }\href@noop
  {} {\bibfield  {journal} {\bibinfo  {journal} {arXiv:0811.1650v3}\ }
  (\bibinfo {year} {2010})}\BibitemShut {NoStop}%
\bibitem [{\citenamefont {Ku}\ \emph {et~al.}(2009)\citenamefont {Ku},
  \citenamefont {Braun},\ and\ \citenamefont {Schwenk}}]{Ku09}%
  \BibitemOpen
  \bibfield  {author} {\bibinfo {author} {\bibfnamefont {M.}~\bibnamefont
  {Ku}}, \bibinfo {author} {\bibfnamefont {J.}~\bibnamefont {Braun}}, and\
  \bibinfo {author} {\bibfnamefont {A.}~\bibnamefont {Schwenk}},\ }\href
  {\doibase 10.1103/PhysRevLett.102.255301} {\bibfield  {journal} {\bibinfo
  {journal} {Phys. Rev. Lett.}\ }\textbf {\bibinfo {volume} {102}},\ \bibinfo
  {pages} {255301} (\bibinfo {year} {2009})}\BibitemShut {NoStop}%
\bibitem [{\citenamefont {Baksmaty}\ \emph {et~al.}(2011)\citenamefont
  {Baksmaty}, \citenamefont {Lu}, \citenamefont {Bolech},\ and\ \citenamefont
  {Pu}}]{Baksmaty11}%
  \BibitemOpen
  \bibfield  {author} {\bibinfo {author} {\bibfnamefont {L.~O.}\ \bibnamefont
  {Baksmaty}}, \bibinfo {author} {\bibfnamefont {H.}~\bibnamefont {Lu}},
  \bibinfo {author} {\bibfnamefont {C.~J.}\ \bibnamefont {Bolech}}, and\
  \bibinfo {author} {\bibfnamefont {H.}~\bibnamefont {Pu}},\ }\href {\doibase
  10.1103/PhysRevA.83.023604} {\bibfield  {journal} {\bibinfo  {journal} {Phys.
  Rev. A}\ }\textbf {\bibinfo {volume} {83}},\ \bibinfo {pages} {023604}
  (\bibinfo {year} {2011})}\BibitemShut {NoStop}%
\bibitem [{\citenamefont {Baur}\ \emph {et~al.}(2009)\citenamefont {Baur},
  \citenamefont {Basu}, \citenamefont {De~Silva},\ and\ \citenamefont
  {Mueller}}]{Baur09}%
  \BibitemOpen
  \bibfield  {author} {\bibinfo {author} {\bibfnamefont {S.~K.}\ \bibnamefont
  {Baur}}, \bibinfo {author} {\bibfnamefont {S.}~\bibnamefont {Basu}}, \bibinfo
  {author} {\bibfnamefont {T.~N.}\ \bibnamefont {De~Silva}}, and\ \bibinfo
  {author} {\bibfnamefont {E.~J.}\ \bibnamefont {Mueller}},\ }\href {\doibase
  10.1103/PhysRevA.79.063628} {\bibfield  {journal} {\bibinfo  {journal} {Phys.
  Rev. A}\ }\textbf {\bibinfo {volume} {79}},\ \bibinfo {pages} {063628}
  (\bibinfo {year} {2009})}\BibitemShut {NoStop}%
\bibitem [{\citenamefont {Diederix}\ and\ \citenamefont
  {Stoof}(2011)}]{Diederix11}%
  \BibitemOpen
  \bibfield  {author} {\bibinfo {author} {\bibfnamefont {J.}~\bibnamefont
  {Diederix}}\ and\ \bibinfo {author} {\bibfnamefont {H.~T.~C.}\ \bibnamefont
  {Stoof}},\ }\href {http://arxiv.org/abs/1102.3320} {\bibfield  {journal}
  {\bibinfo  {journal} {arXiv:1102.3320}\ } (\bibinfo {year}
  {2011})}\BibitemShut {NoStop}%
\bibitem [{\citenamefont {Nascimb\`{e}ne}\ \emph {et~al.}(2009)\citenamefont
  {Nascimb\`{e}ne}, \citenamefont {Navon}, \citenamefont {Jiang}, \citenamefont
  {Tarruell}, \citenamefont {Teichmann}, \citenamefont {McKeever},
  \citenamefont {Chevy},\ and\ \citenamefont {Salomon}}]{Nascimbene09}%
  \BibitemOpen
  \bibfield  {author} {\bibinfo {author} {\bibfnamefont {S.}~\bibnamefont
  {Nascimb\`{e}ne}}\ \emph {et~al.}, }\href {\doibase
  10.1103/PhysRevLett.103.170402} {\bibfield  {journal} {\bibinfo  {journal}
  {Phys. Rev. Lett.}\ }\textbf {\bibinfo {volume} {103}},\ \bibinfo {pages}
  {170402} (\bibinfo {year} {2009})}\BibitemShut {NoStop}%
\bibitem [{\citenamefont {Parish}\ and\ \citenamefont {Huse}(2009)}]{Parish09}%
  \BibitemOpen
  \bibfield  {author} {\bibinfo {author} {\bibfnamefont {M.~M.}\ \bibnamefont
  {Parish}}\ and\ \bibinfo {author} {\bibfnamefont {D.~A.}\ \bibnamefont
  {Huse}},\ }\href {\doibase 10.1103/PhysRevA.80.063605} {\bibfield  {journal}
  {\bibinfo  {journal} {Phys. Rev. A}\ }\textbf {\bibinfo {volume} {80}},\
  \bibinfo {pages} {063605} (\bibinfo {year} {2009})}\BibitemShut {NoStop}%
\bibitem [{\citenamefont {Van~Schaeybroeck}\ and\ \citenamefont
  {Lazarides}(2007)}]{Schaeybroeck07}%
  \BibitemOpen
  \bibfield  {author} {\bibinfo {author} {\bibfnamefont {B.}~\bibnamefont
  {Van~Schaeybroeck}}\ and\ \bibinfo {author} {\bibfnamefont {A.}~\bibnamefont
  {Lazarides}},\ }\href {\doibase 10.1103/PhysRevLett.98.170402} {\bibfield
  {journal} {\bibinfo  {journal} {Phys. Rev. Lett.}\ }\textbf {\bibinfo
  {volume} {98}},\ \bibinfo {pages} {170402} (\bibinfo {year}
  {2007})}\BibitemShut {NoStop};  
  \bibfield  {author} {\bibinfo {author} {\bibfnamefont {B.}~\bibnamefont
  {Van~Schaeybroeck}}\ and\ \bibinfo {author} {\bibfnamefont {A.}~\bibnamefont
  {Lazarides}},\ }\href {\doibase 10.1103/PhysRevA.79.053612} {\bibfield
  {journal} {\bibinfo  {journal} {Phys. Rev. A}\ }\textbf {\bibinfo {volume}
  {79}},\ \bibinfo {pages} {053612} (\bibinfo {year} {2009})}\BibitemShut
  {NoStop}%
\bibitem [{\citenamefont {Luo}\ and\ \citenamefont {Thomas}(2009)}]{Luo09}%
  \BibitemOpen
  \bibfield  {author} {\bibinfo {author} {\bibfnamefont {L.}~\bibnamefont
  {Luo}}\ and\ \bibinfo {author} {\bibfnamefont {J.~E.}\ \bibnamefont
  {Thomas}},\ }\href
  {http://www.springerlink.com/content/77p2u304vk686080/fulltext.pdf}
  {\bibfield  {journal} {\bibinfo  {journal} {J. Low Temp. Phys.}\ }\textbf
  {\bibinfo {volume} {154}},\ \bibinfo {pages} {1} (\bibinfo {year}
  {2008})}\BibitemShut {NoStop}%
\bibitem [{\citenamefont {Hu}\ \emph {et~al.}(2010)\citenamefont {Hu},
  \citenamefont {Liu},\ and\ \citenamefont {Drummond}}]{Hu10}%
  \BibitemOpen
  \bibfield  {author} {\bibinfo {author} {\bibfnamefont {H.}~\bibnamefont
  {Hu}}, \bibinfo {author} {\bibfnamefont {X.-J.}\ \bibnamefont {Liu}}, and\
  \bibinfo {author} {\bibfnamefont {P.~D.}\ \bibnamefont {Drummond}},\
  }\href@noop {} {\bibfield  {journal} {\bibinfo  {journal} {New J. Phys.}\
  }\textbf {\bibinfo {volume} {12}},\ \bibinfo {pages} {063038} (\bibinfo
  {year} {2010})}\BibitemShut {NoStop}%
\bibitem [{\citenamefont {Nascimb\`{e}ne}\ \emph {et~al.}(2010)\citenamefont
  {Nascimb\`{e}ne}, \citenamefont {Navon}, \citenamefont {Jiang}, \citenamefont
  {Chevy},\ and\ \citenamefont {Salomon}}]{Nascimbene10}%
  \BibitemOpen
  \bibfield  {author} {\bibinfo {author} {\bibfnamefont {S.}~\bibnamefont
  {Nascimb\`{e}ne}}, \bibinfo {author} {\bibfnamefont {N.}~\bibnamefont
  {Navon}}, \bibinfo {author} {\bibfnamefont {K.~J.}\ \bibnamefont {Jiang}},
  \bibinfo {author} {\bibfnamefont {F.}~\bibnamefont {Chevy}}, and\ \bibinfo
  {author} {\bibfnamefont {C.}~\bibnamefont {Salomon}},\ }\href@noop {}
  {\bibfield  {journal} {\bibinfo  {journal} {Nature}\ }\textbf {\bibinfo
  {volume} {463}},\ \bibinfo {pages} {1057} (\bibinfo {year}
  {2010})}\BibitemShut {NoStop}%
\bibitem [{\citenamefont {Sommer}\ \emph
  {et~al.}(2011{\natexlab{b}})\citenamefont {Sommer}, \citenamefont {Ku},\ and\
  \citenamefont {Zwierlein}}]{Sommer11b}%
  \BibitemOpen
  \bibfield  {author} {\bibinfo {author} {\bibfnamefont {A.}~\bibnamefont
  {Sommer}}, \bibinfo {author} {\bibfnamefont {M.}~\bibnamefont {Ku}}, and\
  \bibinfo {author} {\bibfnamefont {M.~W.}\ \bibnamefont {Zwierlein}},\ }\href
  {\doibase http://dx.doi.org/10.1088/1367-2630/13/5/055009} {\bibfield
  {journal} {\bibinfo  {journal} {New J. Phys.}\ }\textbf {\bibinfo {volume}
  {13}},\ \bibinfo {pages} {055009} (\bibinfo {year}
  {2011}{\natexlab{b}})}\BibitemShut {NoStop}%
\bibitem [{\citenamefont {Sommer}\ \emph
  {et~al.}(2011{\natexlab{a}})\citenamefont {Sommer}, \citenamefont {Ku},
  \citenamefont {Roati},\ and\ \citenamefont {Zwierlein}}]{Sommer11a}%
  \BibitemOpen
  \bibfield  {author} {\bibinfo {author} {\bibfnamefont {A.}~\bibnamefont
  {Sommer}}, \bibinfo {author} {\bibfnamefont {M.}~\bibnamefont {Ku}}, \bibinfo
  {author} {\bibfnamefont {G.}~\bibnamefont {Roati}}, and\ \bibinfo {author}
  {\bibfnamefont {M.~W.}\ \bibnamefont {Zwierlein}},\ }\href {\doibase
  doi:10.1038/nature09989} {\bibfield  {journal} {\bibinfo  {journal} {Nature}\
  }\textbf {\bibinfo {volume} {472}},\ \bibinfo {pages} {201} (\bibinfo {year}
  {2011}{\natexlab{a}})}\BibitemShut {NoStop}%
\end{thebibliography}

%

\end{document}